\documentclass[sigconf]{acmart}

\usepackage{enumitem}
\usepackage{booktabs}
\usepackage{graphicx}
\usepackage{subcaption}

\copyrightyear{2022} 
\acmYear{2022} 
\setcopyright{rightsretained} 
\acmConference[UIST '22 Adjunct]{The Adjunct Publication of the 35th Annual ACM Symposium on User Interface Software and Technology}{October 29-November 2, 2022}{Bend, OR, USA}
\acmBooktitle{The Adjunct Publication of the 35th Annual ACM Symposium on User Interface Software and Technology (UIST '22 Adjunct), October 29-November 2, 2022, Bend, OR, USA}
\acmDOI{10.1145/3526114.3558698}
\acmISBN{978-1-4503-9321-8/22/10}

\begin{document}

\title[Understanding Concerns of Real-World Users on Smart Mobile Health Applications]{``\textit{Inconsistent Performance}'': Understanding Concerns of Real-World Users on Smart Mobile Health Applications Through Analyzing App Reviews}

\author{Banafsheh Mohajeri}
\email{banafsheh.mohajeri@polymtl.ca}
\affiliation{%
  \institution{Polytechnique Montreal}
  \city{Montreal}
  \state{QC}
  \country{Canada}
}

\author{Jinghui Cheng}
\email{jinghui.cheng@polymtl.ca}
\affiliation{%
  \institution{Polytechnique Montreal}
  \city{Montreal}
  \state{QC}
  \country{Canada}
}

\begin{abstract}
While smart mobile health apps that adapt to users' progressive individual needs are proliferating, many of them struggle to fulfill their promises due to an inferior user experience. Understanding the concerns of real-world users related to those apps, and their smart components in particular, could help advance the app design to attract and retain users. In this paper, we target this issue through a preliminary thematic analysis of 120 user reviews of six smart health apps. We found that accuracy, customizability, and convenience of data input are primary concerns raised in real-world user reviews. Many concerns on the smart components are related to the trust issue of the users towards the apps. However, several important aspects such as privacy and fairness were rarely discussed in the reviews. Overall, our study provides insights that can inspire further investigations to support the design of smart mobile health apps. 
\end{abstract}

\ccsdesc[500]{Human-centered computing~Empirical studies in HCI}
\ccsdesc[500]{Applied computing~Health care information systems}

\keywords{Mobile health, smart component, app review}

\maketitle

\section{Introduction}
Mobile health apps are increasingly popular. They help individuals with numerous types of health-related objectives such as weight loss, self-management of health problems, and self-diagnosis, to just name a few. With the growth of the health apps market, there are more and more apps integrating some form of smart components, leveraging techniques such as machine learning, in order to offer capabilities of  progressively adapting to individual user's needs.

However, preliminary research demonstrated that many users abandon health apps after the initial enthusiasm~\cite{mustafa2022}, in part because the users do not possess confidence in the apps' ability and that they are not satisfied with the user experience that they receive. In particular, smart health apps, although showing great potentials, can present new and unique challenges related to the user experience design. These challenges are originated partly from the algorithms that enable the smart capacity, whose performance is not always accurate or even predictable~\cite{Dove2017}. If not communicated appropriately, this problem can discourage the users from continuously using the app. On the other hand, a well designed app conveying the adequate information about the smart component, as well as its corresponding capabilities, can enhance the user experience. Sufficient user control and feedback can also help improve the abilities and performance of the smart component~\cite{Amershi2019}.

There has been previous studies aimed at (1) understanding user perceptions and opinions towards apps for a specific health purpose (e.g.,~\cite{Niess2021}) and (2) reviewing the landscape of different kinds of commercially available health apps (e.g.,~\cite{kao2017, Lau2020}). However, little is known about the concerns from the perspectives of \textit{real-world users} who interact with smart mobile health apps on a daily basis, particularly with the smart components that are increasingly prevalent in those apps. This knowledge is important in advancing the design, promoting the adoption, and thus fulfilling the potentials and promises of smart health apps.

In this paper, we address this gap by analyzing the users reviews of popular commercially available smart mobile health apps. Previous studies have used app user reviews as a proxy to understand users' concerns in various types of apps~\cite{Ghosh2018,Prakash2020,Phillips2021}. Through this analysis, we harvest the overall concerns from real-world users about the design decisions and capabilities of the apps affecting the user experience, especially those related to the smart component. 

In order to reach this research goal, we conducted a thematic analysis of user reviews to six popular mobile health apps that have smart components. Our results indicated that real-world users especially cared about the apps' accuracy and customizability. Many factors of users' opinions are related to the trust issue. However, some important aspects such as privacy and fairness are rarely discussed in the user reviews. Our results support future research focused on advancing the user experience of smart mobile apps.
\section{Methods}
We sampled popular health mobile applications (apps) that included a smart component and analyzed the user reviews of these apps. We focused on Android apps on the Google Play store because it is considered as the most popular mobile platform~\cite{MobileOperatingSystem}. We sampled apps according to the following steps. First, the apps are selected from the \textit{Health \& Fitness} and the \textit{Medical} categories of the Google Play store. Second, we collected the apps that had more than 10 million installs to ensure that the analyzed apps had a large user base. Third, to avoid homogeneous results we included only one app for each specific health function (e.g., diet coach, step tracking, etc.) and excluded apps developed by the same company. Finally, we inspected the main functionalities of each app by installing them and using them for two weeks, to filter out any apps that did not have a smart component. We considered a component is ``smart'' if it is capable of being personalized and learns from the users' choices and/or usage history. These steps resulted in the following six apps that were the focus of further analysis: (1) Google Fit: Activity Tracking, (2) Six Pack in 30 Days, (3) Headspace: Meditation \& Sleep, (4) YAZIO Fasting \& Food Tracker, (5) WebMD: Symptom Checker, and (6) Step Tracker -- Pedometer.

After selecting the apps, we collected the top 20 user reviews of each app sorted in the \textit{most relevant} mode of Google Play store, resulting in 120 user reviews. We then conducted a thematic analysis~\cite{Vaismoradi2013} of these user reviews to understand users' concerns related to the smart components in the apps. Particularly, the two authors conducted a series of iterative open coding on the user reviews; each review can be applied with multiple codes. Then in several extensive meetings, the authors merged and categorized the codes into higher-level themes with the help of an affinity diagram activity.

\section{Results}

We found that users frequently mentioned the smart components in their reviews. We grouped these concerns in the following categories and present them in the descending order of frequency each category appeared in our sample.


\textbf{Accuracy} (mentioned $N=53$ times).
When users felt that the app is accurate and can produce the correct information, they generally had a better experience. For example, a Step Tracker - Pedometer reviewer noted that they like it because {\itshape``Step counts, time recording, and calories burned are accurate.''} On the flip side, one of the biggest complaints from the users was the lack of accuracy of the smart component. Sometimes the complaint was about how the app miscalculates the data received from the users, such as the number of steps or the walking distance. Other times the complaint was about how the app failed to record user activities completely. As an example of the latter case, a user reviewing GoogleFit mentioned: {\itshape``It will count my steps when I'm standing still but not when I'm actually walking. I just lost 90 minutes of logging because it wouldn't count my steps.''} This inaccuracy also sometimes caused the users to lose their faith in the app and no longer be able to trust it. For example, another GoogleFit reviewer detailed this point of view by saying: {\itshape``Unreliable and inaccurate tracking. Heart points disappear all of a sudden. On the day of workout it shows the heart points, and next day it becomes 0. Not recommended to track your daily.''}

\textbf{Customizability} ($N=38$).
The users appreciated it when the apps support users to make changes to the plan or the configuration of the smart component based on their personal preferences. They wanted not to be stuck with the default recommendation by the algorithms. They generally thought a more personalized plan suited more for their health needs and preferred to have control over the recommendations and suggestions. A Headspace user explained how the meditation sessions are particularly helpful for them in part because {\itshape ``being able to choose how long they last is great.''} Predictably, users did not like apps that were not able to provide them with smart services based on their specific health-related needs. Apps which failed in this regard sometimes would become useless for users as they were not able to achieve their personal health and fitness goals with them. For example, a YAZIO reviewer mentioned they could not really use the app since they do not {\itshape ``see an option to change how much you wanted to lose and how fast.''}

\textbf{Convenience of data input} ($N=33$).
Many users put a great deal of emphasis on how easy it is for them to use the app without having to invest too much time or manual labour. The most prominent problem the users mentioned was that the apps were hard to use and took too much time to just be able to function, to the point that for some users they were not a really viable option anymore. Many occurrences of this problem happened with apps that required the users to fill in very detailed activities and habits, such as food intake or exercise, manually. For example, a YAZIO user noted how they are not very inclined to keep using the app because it is {\itshape``Too cumbersome to actually track food every day''}

\textbf{Health efficacy} ($N=20$). 
Users complimented the apps whenever they had a smart component which was able to reliably perform the intended health-related functions. For example, the users of Headspace noted how it helps them {\itshape ``to get to sleep''} and {\itshape ``concentrate when I need to''} by providing good suggestions, sleepcasts, and programs. Another user of Headspace corroborated the app's ability to help users address multiple mental health issues effectively by saying: {\itshape ``Highly recommend for stress, anxiety and focus! they also address pain, anger management, depression and so much more.''}

\textbf{Transparency and clarity} ($N=13$).
The users generally tended to like the app more when they could figure out the required information and understand how the algorithm works without trouble and could operate the app hustle free. A user explained why they chose Step tracker- Pedometer app by saying how {\itshape``it has easy-to-read charts to help you visualize your progress''}. On the other hand, users tended to complain when they find the app hard to understand, saw the user interface as unclear and confusing, or thought that they had to put a lot of work into understanding it.

\textbf{Adaptability} ($N=9$).
The users appreciated the apps that provided suggestions adapted to the user's past information and behaviors. For example, a Headspace user complimented the app by stating: {\itshape``App is smooth to use and allows you to keep track of your stress, then suggests programs that could help at the moment''}. Similarly, users tend to like apps that can make changes based on the feedback given by the users. A user reviewing the Six Pack in 30 Days app counted its ability to change based on the user’s feedback as a positive point by saying {\itshape``The coach system can adjust the workouts based on the feedback you give.''}

\textbf{Scientific reliability} ($N=6$).
Users trusted apps more when they felt that the plans and services provided for them with the smart component have been built with consultation from health professionals and/or based on scientific data. These factors make the users feel that the app is safe and effective to use. A user reviewing Sixpack in 30 days highlighted this fact by mentioning that they trust the app because they feel like it gave them a {\itshape ``systematic and scientific way of reducing your fat and get a tuned body.''}

\textbf{Response time} ($N=5$).
Users made complaints when the smart component took too long to respond and made the whole app slow. For example, a reviewer mentioned how they are uncertain about continuing to use YAZIO because for them it {\itshape``feels slow''}.

\textbf{Perceived ``smartness''} ($N=4$).
Users liked the apps that felt smart to them. Sometimes they mentioned that they can trust the app's guidance like a real human professional. For example, a user noted how much Six Pack in 30 Days feels like a real coach by saying it feels like {\itshape``You have your virtual trainer with your as if you are in a gymnasium''}. Some users preferred that in case of uncertainty the smart component show them multiple options instead of just one; some were upset when it failed to do so. One user noted this problem when reviewing WebMD by complaining about how when it is not certain {\itshape``it doesn't give you any possible results.''}
\section{Discussion}
Through our analysis, we found that the most frequent concerns on the smart components of these apps from the real-world users' perspectives are \textit{accuracy}, \textit{customizability}, and \textit{convenience of data input}. 
Together, our results offer the following insights.

First, \textbf{users tend to focus more frequently on \textit{customizability} than \textit{adaptability} in the reviews}. In our analysis, we considered \textit{customizability} as the ability of the app to give users the freedom and control of changing the behavior of the smart component, while \textit{adaptability} as the ability of the app to automatically change its own behavior according to user data input. Our results indicate that users seem to pay more attention to the ability of gaining control over the app, indicating their awareness, and perhaps preference, to direct manipulation over full automation. Another possible explanation is that the user experience of \textit{adaptability} is often ``seamless'', thus difficult to pinpoint.

Second, \textbf{there is a noteworthy focus on the trust issue of the smart components in the reviews, reflected in four codes: \textit{accuracy}, \textit{health efficacy}, \textit{transparency and clarity}, and \textit{scientific reliability}.} These codes covered about half of all the codings of the smart component concerns. Jacovi et al.~\cite{Jacovi2021} categorized the causes of trust of a user to an AI system as two types: (1) intrinsic trust that relies on the ``observable decision process'' of the AI component and (2) extrinsic trust that relies on empirical evaluation of performance. Our analysis revealed that users of smart health apps tend to focus on extrinsic trust (i.e., the \textit{accuracy}, \textit{health efficacy}, and \textit{scientific reliability} concerns), rather than intrinsic trust (i.e., the \textit{transparency and clarity} concern). Interestingly, the three extrinsic trust concerns demonstrate a direct mapping to the \textit{Product Performance}, \textit{Satisfaction with Previous Interactions}, and \textit{Reputation} aspects in Johnson and Grayson's model~\cite{JOHNSON2005500} that contributes to the cognitive trust of customers in a service provider.

Third, \textbf{in the user reviews we did not find direct discussion on many other important aspects such as safety, security, privacy, and fairness.} While we do not know exactly why this is the case, there are several possible explanations. It could well be due to the reason that the users do not consider these aspects as important or lack awareness of them. Or, it is because that the reviews only reflect the opinions of the users who are willing to use the apps and have accepted the possible risks related to these aspects. Further study is needed to investigate these factors.

\textbf{Limitations and future work.}
Our study was constricted to only six apps among the many available health apps, so the generalizability of our findings could be limited and future investigations are required. Related, all of the reviews were from Android users; although other mobile platforms share many of the same apps, further research is needed to explore user opinions on other platforms. Further, we only used reviews to gather the user opinions, so we were not able to fully understand the users' perceptions, especially those who did not leave reviews; we plan to address this issue in the future through methods such as interviews and user studies. 

\section{Conclusion}
In this paper, we conducted a thematic analysis on 120 user reviews of six mobile health apps that include a smart component to understand the most prominent concerns from the real-world users' perspectives. Our findings indicate that user reviews of these apps are frequently focused on their smart components; particularly, \textit{accuracy}, \textit{customizability}, and \textit{convenience of data input} are the primary concerns. While many user reviews are focused on the trust relationship between the users and the app, very few touched other aspects such as privacy and fairness. These findings will inspire further investigations towards guidelines that support the design of smart mobile health apps to fully achieve their potentials.

\begin{acks}
This research was partially supported by the Canada Research Chairs Program, as well as the Discovery Grants Program of the Natural Sciences and Engineering Research Council of Canada.
\end{acks}

\bibliographystyle{ACM-Reference-Format}
\bibliography{references}

\end{document}